\documentclass[10]{iopart}
\usepackage[bookmarks = true, pdfpagemode = None, pdfstartview = FitH, colorlinks = false,
urlcolor = blue, backref]{hyperref} 
\usepackage[]{cite}
\usepackage{graphicx}
\usepackage{url}

\usepackage{amssymb}

\newtheorem{definition}{Defintion}
\newtheorem{lemma}{Lemma}
\newtheorem{proof}{Proof:}
\newtheorem{theorem}{Theorem}
\newtheorem{conjecture}{Conjecture}

\begin{document}

\title[Yang, et. al.]{
Realizing Ternary Quantum Switching Networks without Ancilla Bits}

\author{Guowu Yang,\footnote[1]{Guowu Yang is the author with whom electronic correspondence shall be
addressed:~\href{mailto:guowu@ece.pdx.edu}{\tt guowu@ece.pdx.edu}}
 Xiaoyu Song and Marek Perkowski
 }\address{Department of Electrical \& Computer Engineering,
Portland State University, 1900 SW Fourth Avenue, P.O. Box 751,
Portland, Oregon 97201, USA. }

\author{Jinzhao Wu}\address{Fakult\"{a}t fur Mathematik und Informatik, Universit\"{a}t
Mannheim, Mannheim, Germany.}

\begin{abstract}
This paper investigates the synthesis of quantum networks built to
realize ternary switching circuits in the absence of ancilla bits.
The results we established are twofold.  The first shows that
ternary Swap, ternary Not and ternary Toffoli gates are universal
for the realization of arbitrary $n\times n$ ternary quantum
switching networks without ancilla bits. The second result proves
that all $n\times n$ quantum ternary networks can be generated by
Not, Controlled-Not, Multiply-Two, and Toffoli gates. Our approach
is constructive.
\end{abstract}


\submitto{\JPA}

\maketitle

\section{Introduction}
\textbf{Q}uantum computation connects ideas from computer science
and physics~\cite{NielsenC:00}.  Reversible circuits are a necessary
subclass whose realization is required for any quantum computer to
be universal. Three state quantum systems have recently been
discussed in the framework of cryptography~\cite{Pasquinucci:00},
and the concept of a qudit cluster state has been
proposed~\cite{Zhou:03}. Qudit systems received further study
in~\cite{DaboulYS:03} and~\cite{Kunio} wherein quantum hybrid gates
acting on tensor products of qudits of different dimensions were
discussed.  Recently synthesis for d-level systems showing
asymptotic optimality was also proposed~\cite{BullOB:AOQCFDS:05}.
The study in~\cite{DaboulYS:03} and~\cite{MuthukrishnanS:MVQG:00}
found hybrid quantum gates that, when considered to be controlled by
and act on three level quantum systems define the hybrid Toffoli,
Swap and Not gates used in this paper. The physical realization of
these hybrid gates might be accomplished via spin
systems~\cite{DaboulYS:03, BartlettGS:QESHO:02} or quantum harmonic
oscillators~\cite{DaboulYS:03, BartlettGS:QESHO:02}. A universal set
of ternary quantum gates enables the realization of any tristate
switching network on a candidate qudit realization.

The computer science community has also experienced recent interest
in the universal sets of gates required for ternary quantum
computing systems; the main results of which appear
in~\cite{MillerMD:05, KhanPKK:TGFSOP:2004, Rabadi:MVGF:2001}. In
these gates, arbitrary Galois field operations are used in the
so-called Toffoli gates of the ESOP-based (Exclusive
Sums-of-Products) realization of binary reversible circuits, where
Galois addition and multiplication replace the XOR and And gates,
respectively. The ESOP circuit synthesis programs use heuristic
rule-based search strategy to minimize each output as an Exclusive
Sum of Products realized as $k$-input ($k\leq n$) Toffoli gates. We
observe that the universality discussed in the literature has an
assumption that the inputs of gates can be set to constant values,
thus ancilla bits are used~\cite{MillerMD:05, KhanPKK:TGFSOP:2004,
Rabadi:MVGF:2001}. These programs can be applied to large functions
but their disadvantage is that they create $m$ ancilla bits (one for
each output) and use multi-input gates that may be expensive.
Although~\cite{DaboulYS:03} discussed entanglement
generation\footnote{The reader wishing to develop background in the
theory of quantum computation should consult the textbook by Nielsen
and Chuang~\cite{NielsenC:00} and the references therein.} with and
without ancilla qudits, in both the physics and computer science
community neither the ternary switching universality of the
introduced sets of gates nor the proof of a synthesis algorithms
convergence was given.

Group theory~\cite{DixonM:PG} has found particular use to generate
reversible logic circuits~\cite{DeVosRS:GGRLG:99}.  Some notable
results appear in~\cite{DeVosRS:GGRLG:99}
and~\cite{Storme:GTAORLG:99}\cite{SongYPW:ACORLG:05}\cite{YangHSP:MBRLG:05}
and are applicable to the synthesis of quantum switching networks.
The motivation of this paper is to find the universality of a gate
family~\cite{Toffoli:BEOICF:81} to be used in synthesis of ternary
reversible circuits without ancilla bits. We prove that ternary
Swap, Not and Toffoli gates~\cite{DaboulYS:03} are universal for
realization of arbitrary ternary $n\times n$ reversible circuits
without ancilla bits. Moreover we create an algorithm for one of
these gate families that is provably convergent. Our algorithm is
constructive and effective in both space and time resources.

This paper is organized as follows. First, in Sec.~\ref{sec:main},
we introduce some basic definitions of ternary switching networks
and the needed group theory natation, terms and results. We then
present our main results: Theorem~\ref{theorem:generate}
and~\ref{theorem:generate2} after four Lemmas. Second, we conclude
this paper. Finally, in the appendix, we prove
Lemma~\ref{lemma:3defvec}, and present two examples to illustrate
the synthesis process for a given ternary reversible circuit.

\section{Main results}\label{sec:main}

This section begins by presenting some basic definitions of ternary
switching networks and the needed group theory notation and terms.

\begin{definition}[Ternary reversible gate]
Let $B = \{0, 1, 2\}$. A ternary logic circuit $f$ with $n$ input
variables, $B_1,\ldots,B_n$, and $n$ output variables, $P_1,\ldots,
P_n$, is denoted by $f: B^n\rightarrow{}B^n$, where $<B_1,\ldots,
B_n>\in{}B^n$ is the input vector and $<P_1,\ldots,P_n>\in{}B^n$ is
the output vector. There are $3^n$ different assignments for the
input vectors. A ternary logic circuit $f$ is reversible if it is a
one-to-one and onto function (bijection). A ternary reversible logic
circuit with $n$ inputs and $n$ outputs is also called an $n\times
n$ ternary reversible gate. There are a total of $(3^n)!$ different
{$n\times n$} ternary reversible circuits.
\end{definition}
The concept of a permutation group and its relationship with
reversible circuits will now be introduced.

\begin{definition}[Permutation]
Let $M = \{d_1, d_2,\ldots,d_k\}$. A bijection~\footnote{Bijection:
one-to-one, and onto mapping.} of $M$ onto itself is called a
permutation on $M$. The set of all permutations on $M$ forms a group
under composition of mappings, called a symmetric group on $M$. It
is denoted by $S_k$~\cite{DixonM:PG}. A permutation group is simply
a subgroup~\cite{DixonM:PG} of a symmetric group.
\end{definition}
A mapping $s: M\rightarrow M$ can be written as:
\begin{equation}
s =\Big({d_1,d_2,\ldots,d_k \atop
d_{i_1},d_{i_2},\ldots,d_{i_1}}\Big)
\end{equation}
Here we use a product of disjoint cycles as an alternative notation
for a mapping~\cite{DixonM:PG}. For example,
\begin{equation}
\Big({d_1,d_2,d_3,d_4,d_5,d_6,d_7,d_8,d_9 \atop
d_1,d_4,d_7,d_2,d_5,d_8,d_3,d_6,d_9}\Big)
\end{equation}
can be written as $(d_2, d_4)(d_3, d_7)(d_6, d_8)$. Denote "( )" as
the identity mappings direct wiring and call this the unity element
in a permutation group. The inverse mapping of mapping $s$ is
denoted as $s^{-1}$. As per convention, a product $s\star t$ of two
permutations applies mapping $s$ before $t$.

We order the $3^n$ different $n$-input assignment vectors as:
\begin{equation}
\hspace{-1cm} (0,0,\ldots,0), (1,0,\ldots,0), (2,0,\ldots,0),
(0,1,\ldots,0),\ldots, (2,2,\ldots,2),
\end{equation} and denote them by $a_1, a_2,
a_3$,\ldots, $a_m$, where $m = 3^n$. Thus a $n\times n$ ternary
reversible circuit is just a permutation in $S_m$ (\emph{where $m =
3^n$}), and vice versa. Cascading two gates is equivalent to
multiplying two permutations. In what follows, no distinction
between an $n\times n$ reversible gate and a permutation in $S_m$
(\emph{where $m = 3^n$}) will be made.

\begin{definition}[Swap Gate]
A Swap gate $E_{i,j}$ exchanges the $i^{th}$ bit $B_i$ and the
$j^{th}$ bit $B_j$, i.e. $P_i=B_j, P_j=B_i; P_r=B_r,$ if $r\neq
i,j$.
\end{definition}

\begin{definition}[Ternary Not Gate]
A Ternary Not Gate $N_j$ is defined as: $P_j = B_j\bigoplus
_31$\footnote{$\bigoplus _3$ denotes addition modulo 3}; $\quad P_i
= B_i,\quad if \quad i\neq j.\quad 1\leq j\leq n.$
\end{definition}

\begin{definition}[Ternary Toffoli Gate]
A Ternary Toffoli Gate T is defined such that if $B_2 = B_3 =\ldots
= B_n = 1$, then $P_1 = B_1 \bigoplus _31$; otherwise, $P_1 = B_1$,
whereas $P_i = B_i$, for $i\neq 1$. In other words, it maps $d_1$ to
$d_2$, $d_2$ to $d_3$, $d_3$ to $d_1$, respectively, where
$d_1=(0,1,1,\ldots,1), d_2 = (1,1,1,\ldots,1), d_3 =
(2,1,1,\ldots,1),$ and the other assignment vectors do not change.
\begin{equation}
\left[\begin{array}{c}d_1 \\ d_2 \\ d_3 \\ \ldots
\end{array}\right] = \left[\begin{array}{c}0,1,1,\ldots,1\\
1,1,1,\ldots,1\\ 2,1,1,\ldots,1\\ others\end{array}\right] {T
\atop \rightarrow} \left[\begin{array}{c}1,1,1,\ldots,1\\
2,1,1,\ldots,1\\ 0,1,1,\ldots,1\\ others\end{array} \right] =
\left[\begin{array}{c}d_2\\ d_3\\ d_1\\ \ldots \end{array}\right]
\end{equation}
\end{definition}
From the definition of $T$, we have $T = (d_1, d_2, d_3)$. Thus, $T$
is a 3-cycle, and $T^{-1} = T\star T$, $(T\star T)^{-1} = T$.

\begin{definition}[j-cycle]
Let $S_k$ be a symmetric group of symbols \{$d_1, d_2, \ldots,
d_k$\}, then $(d_{i_1}, d_{i_2}, \ldots, d_{i_j})$, where $j\leq k$,
is called a $j$-cycle. In particular, a $j$-cycle $(d_i, d_{i+1},
\ldots, d_{i+j-1})$ is called a neighbor $j$-cycle of $S_k$, for
$\forall 1 \leq i \leq k-j+1$
\end{definition}

\begin{definition}[even permutation and odd permutation]
A permutation is even if it is a product of an even number of
2-cycles and odd if it is an odd a product number of 2-cycles.
\end{definition}
Obviously, a 3-cycle is an even permutation. For instance, (1, 3, 2)
= (2, 3) (3, 1). The product of some even permutations is also an
even permutation. The product of an odd number of odd permutations
is an odd permutation. The product of an even number of even
permutations with an odd number of odd permutations is an odd
permutation. The product of an even number of odd permutations is an
even permutation.

\begin{lemma}\label{lemma:Eij}
$E_{i,j}$ is: a product of $3^{n-1}$ disjoint 2-cycle permutations,
an odd permutation and $(E_{i,j})^{-1} = E_{i,j}$.
\end{lemma}
\begin{proof}From the definition of $E_{i,j}$, we have the mapping of
$E_{i,j}$ in (\ref{eqn:map Eij}), thus the disjoint 2-cycle's $(b_1,
b_2)$, $(b_3, b_4)$, $(b_5, b_6)$ are in $E_{i,j}$. There are
$3^{n-2}$ cases for the assignments of the $n-2$ positions except
$B_i$ and $B_j$. Thus, there are $3^{n-2}\star 3 = 3^{n-1}$ disjoint
2-cycles in $E_{i,j}$. The other vectors do not change. Therefore,
$E_{i,j}$ is a product of these $3^{n-1}$ disjoint 2-cycles. So
$E_{i,j}$ is an odd permutation and $(E_{i,j})^{-1} = E_{i,j}$.  For
example, when $n = 2$, we have $E_{1,2} = (d_2, d_4)(d_3, d_7)(d_6,
d_8)$.
\begin{equation}\label{eqn:map Eij}
\hspace{-2.5cm}
\begin{array}{@{}c@{}c@{}c@{}c@{}c@{}c@{}c@{}}
&&1,\ldots,i,\ldots,j,\ldots,n&& 1,\ldots,i,\ldots,j,\ldots,n&&\\
\left[\begin{array}{c} b_1\\ b_2\\ b_3\\ b_4\\ b_5\\ b_6\\ \ldots
\end{array}\right]
&=
&\left[\begin{array}{c} B_1,\ldots,1,\ldots,0,\ldots,B_n\\
B_1,\ldots,0,\ldots,1,\ldots,B_n\\
B_1,\ldots,1,\ldots,2,\ldots,B_n\\
B_1,\ldots,2,\ldots,1,\ldots,B_n\\
B_1,\ldots,2,\ldots,0,\ldots,B_n\\
B_1,\ldots,0,\ldots,2,\ldots,B_n\\
other vectors\end{array}\right] &{E_{i,j} \atop \rightarrow}
&\left[\begin{array}{c} B_1,\ldots,0,\ldots,1,\ldots,B_n\\
B_1,\ldots,1,\ldots,0,\ldots,B_n\\
B_1,\ldots,2,\ldots,1,\ldots,B_n\\
B_1,\ldots,1,\ldots,2,\ldots,B_n\\
B_1,\ldots,0,\ldots,2,\ldots,B_n\\
B_1,\ldots,2,\ldots,0,\ldots,B_n\\
other vectors\end{array}\right] &=
&\left[\begin{array}{c}b_2\\
b_1\\ b_4\\ b_3\\ b_6\\ b_5\\ \ldots
\end{array}\right]
\end{array}
\end{equation}
The Proof of Lemma~\ref{lemma:Eij} is therefore complete.
\hfill{$Q.E.D.$}
\end{proof}

\begin{lemma}
$N_i$ is a product of $3^{n-1}$ disjoint 3-cycle permutations and
$(N_i)^{-1} = N_i\star N_i$, $(N_i\star N_i)^{-1} = N_i$.
\end{lemma}
\begin{proof}
    The proof follows similarly to the proof of
    Lemma~\ref{lemma:Eij}. \hfill{$Q.E.D.$}
\end{proof}

\begin{lemma}\label{evenodd}
Let $S_k$ be a symmetric group of letters \{$d_1, d_2, \ldots,
d_k$\}. Then every even permutation can be generated by using only
neighbor 3-cycles. Obviously, every even permutation can be also
generated by using only 3-cycles.
\end{lemma}
Lemma~\ref{evenodd} is a well-known result in permutation group
theory. It can be found in many textbooks such as~\cite{DixonM:PG}.

\begin{lemma}\label{lemma:3defvec}
For any three different assignment vectors $u$, $s$ and $t$, the
3-cycle permutation $(u, s, t)$ can be generated by Not gate $N_j$,
Swap gate $E_{i,j}$, and Toffoli gate T.
\end{lemma}
The proof of Lemma~\ref{lemma:3defvec} and some examples
illustrating the synthesis process for a given ternary reversible
circuit are given in Appendix.


\begin{theorem}\label{theorem:generate}
All $n\times n$ ternary reversible circuits can be generated by
Swap, Not, and Toffoli gates.
\end{theorem}

\begin{proof}
Let $g$ be a $n\times n$ ternary reversible circuit.\\

\textbf{Case 1}: $g$ is an even reversible circuit. According to
Lemma~\ref{evenodd}, $g$ can be generated by some 3-cycle's.
According to Lemma~\ref{lemma:3defvec}, all 3-cycle's can be
generated by Swap, Not, and Toffoli gates. Therefore, $g$ can be
generated by Swap, Not, and Toffoli gates.

\textbf{Case 2}: $g$ is an odd reversible circuit. Then
$E_{1,2}\star g$ is an even reversible circuit. From case 1,
$E_{1,2}\star g$ can be generated by Swap, Not, and Toffoli gates.
$(E_{1,2})^{-1}= E_{1,2}$. Thus, $g$ can be generated by Swap, Not,
and Toffoli gates. \hfill{$Q.E.D.$}
\end{proof}
The following algorithm is given to synthesize any $n\times n$
ternary reversible circuit:\\

\textbf{Algorithm}: Synthesize any $n\times n$ ternary reversible
circuit $g$.

\textbf{Input}: Swap gate, Not gate, Toffoli gate, and $g$;
\begin{enumerate}
\item If $g$ is an even permutation, \\
     then $g = C_1 \star  C_2 \star  \ldots \star  C_s$; $(C_i$ are 3-cycles for $i=1,\ldots{},s)$
\item $C_i = L_{i,1} \star  L_{i,2} \star \ldots \star L_{i,t_i}$; for
$i=1,2,\ldots{},s$. $(L_{i,j}$ are Swap, or Not, or Toffoli gates$)$
\item Return $g = [L_{1,1} \star \ldots \star L_{1,t_1}] \star \ldots
\star [L_{s,1}\star \ldots \star L_{s,t_s}]$. \item If $g$ is an odd
permutation,
     then $E_{1,2}\star g = L_1 \star L_2 \star \ldots \star L_h$;
     $($where $L_i$ are Swap, or Not, or Toffoli gates$)$
\item Return $g = E_{1,2} \star L_1 \star \ldots \star L_h$.
\end{enumerate}

This algorithm can be implemented in terms of the above Lemmas. Line
1 is based on Lemma~\ref{evenodd}. Line 2 is a logical consequence
from Lemma~\ref{lemma:3defvec}. Line 3 is a direct result from line
1 and 2. In terms of Lemma~\ref{lemma:Eij} and lines 1, 2, and 3, we
have Line 4. From line 4 and Lemma~\ref{lemma:Eij}, line 5 is
derived.

In binary reversible logic, there is a result stating that:
\emph{"All $n\times n$ binary reversible circuits can be generated
by Swap, Not, and Toffoli gates~\cite{SongYPW:ACORLG:05,
Toffoli:BEOICF:81}"}.  This leads to Conjecture~\ref{con:allgen}
which represents an open problem. Although it has not been proven
yet, we strongly believe that it is true.

\begin{conjecture}\label{con:allgen}
All $n\times n$ $p$-value $(p\geq 3)$ reversible circuits can be
generated by Swap, Not, and Toffoli gates (change modulo 3 to modulo
$p$).
\end{conjecture}
In the following, we give some properties of other ternary gates.

\begin{definition}[Ternary Controlled-Not Gate]
A Ternary Controlled-Not Gate $C_{j,i}$ is defined as: $P_j =
B_j\bigoplus _3 1$ if $B_i$ = 1, otherwise, $P_j = B_j$; further:
$P_m = B_m$, if $m\neq j$. Where $1\leq j\neq i\leq n$.
\end{definition}

\begin{definition}[Ternary Multiply-Two Gate]
A Ternary Multiply-Two Gate $MT_i$ is defined as: $P_i =
B_i\bigotimes _3 2$; $P_m = B_m$, if $m\neq i$, where $\bigotimes
_3$ is the operation of multiplication by modulo 3. $1\leq i\leq n$.
\end{definition}

\begin{theorem}\label{theorem:generate2}
All $n\times n$ ternary reversible circuits can be generated by Not,
Controlled-Not, Multiply-Two, and Toffoli gates.
\end{theorem}
\begin{proof}
Using algorithm MLR in ~\cite{YangSHP:FSMRC:05}, we obtain:
$$E_{i,j}
= MT_i \star C_{j,i}\star C_{i,j}\star C_{i,j}\star MT_j \star
C_{i,j}\star C_{j,i}\star C_{j,i} \star MT_i \star C_{j,i}\star
C_{i,j}\star C_{i,j}.$$
From Theorem ~\ref{theorem:generate}, we can
draw the conclusion that all $n\times n$ ternary reversible circuits
can be generated Not, Controlled-Not, Multiply-Two, and Toffoli
gates. \hfill{$Q.E.D.$}
\end{proof}
Based on the similarity to binary quantum switching networks, the
set of Not, Controlled-Not, Multiply-Two, and Toffoli gates is a
more practical set for synthesis. C-Not is a known gate and widely
used gate as is the Not gate. The Toffoli is a natural extension of
C-Not and Not gates. Multiply-two is a single qudit gate so it
should be not expensive. The cost of quantum gates dependents on
different technologies. We hope this set has some cost advantage
when it is used to realize any ternary reversible circuit. In this
paper, we just prove that this set is a universal set. But the
synthesis method based on the proof of theorem 2 is not length
efficient. We are still looking for a length efficient synthesis
algorithm with this set.

\section{Conclusion}
We demonstrated that ternary Swap, ternary Not and ternary Toffoli
gates are universal for realization of arbitrary ternary $n\times n$
reversible circuits without ancilla bits. We also proved that all
$n\times n$ ternary reversible circuits can be generated Not,
Controlled-Not, Multiply-Two, and Toffoli gates. Our approach is
constructive, so it is effective in both space and time resources
but not optimal.

The construction of qudit quantum gates (including ternary
reversible gates) was discussed in [5-8]. The costs of multi-level
reversible gates dependents on the realization of technologies. Our
next plan is to find the cost of these ternary reversible gates, and
create an algorithm with optimal cost by using these gates.

\ack We thank Mr. Jacob Biamonte for useful discussions.

\appendix
\section* {Appendix: A proof of Lemma 4}\label{appendix1}
\textbf{Lemma 4}: For any three different assignment vectors $u$,
$s$ and $t$, the 3-cycle permutation $(u, s, t)$ can be generated by
Not gate $N_j$, Swap gate $E_{i,j}$, and Toffoli gate $T$.

\textbf{Proof}: We denote the vectors $u$, $s$ and $t$ as the
following matrix:

\[
P=\left[\begin{array}{c}u \\ s \\ t\end{array}\right] =
\left[\begin{array}{c}u_1,u_2,\ldots,u_n\\
s_1,s_2,\ldots,s_n\\
t_1,t_2,\ldots,t_n
\end{array}
\right]
\]

In the 3-row matrix $P$, a column having different elements is
called a heterogeneous column. Otherwise, it is called homogeneous
column.

Let $H=\left[P \atop Q\right]$  be the matrix composed of all the
$3^n$ different $n$-input assignments where $Q$ is composed of
$3^{n-3}$ different $n$-input assignment vectors except $u$, $s$ and
$t$.

From the definition, the operations of Swap, Not, and Toffoli gates
on $H$ are as follows.

\begin{itemize}
\item Swap gate $E_{i,j}$ interchanges column $i$ and column $j$.
\item Not gate $N_i$ is an operation $\bigoplus_ 3$1 for all
elements in column $i$. \item Toffoli gate $T$ interchanges three
rows: (0,1,1,\ldots,1) to (1,1,1,\ldots,1), (1,1,1,\ldots,1) to
(2,1,1,\ldots,1), (2,1,1,\ldots,1) to (0,1,1,\ldots,1), and the rest
rows remain fixed.
\end{itemize}
Now we consider the matrix $P$ for the following three cases:\\

\textbf{Case 1}: There is only one heterogeneous column in the
matrix $P$.

\begin{enumerate}

\item We can use a Swap gate $E_{i,j}$ to exchange the
heterogeneous column to the first column position.

\item Using Not gates $N_j$, we can assign all the elements in the
homogeneous columns as values 1.

\item Using Toffoli gate $T$ or $T\star T$ gates (if ($u_1 ,s_1 ,t_1$)
= (0,1,2), or (1,2,0), or (2,0,1), use $T$, otherwise $T\star T$),
we can reorder the rows $r_1, r_2, r_3$ to $r_2, r_3, r_1$ in the
matrix $P$.

\item Finally, using the inverse of the Not and Swap gates used in
steps 2 and 1 to recover the changed digital numbers, we obtain the
3-cycle $(u, s, t)$.
\end{enumerate}

Denote $P^{(i)}$ and $Q^{(i)}$ as the image matrices of $P$ and $Q$
after the $i^{th}$ step, $i=1,2,3,4$. Then the operations of the
$4^{th}$ step are as follows:
\[
P^{(3)}{step 4\atop \rightarrow}P^{(4)}= \left[\begin{array}{c}s\\ u\\
t\end{array}\right], Q^{(3)}{step 4\atop \rightarrow}Q^{(4)}=Q
\]
This process means that an arbitrary 3-cycle permutation $(u, s, t)$
with only one heterogeneous column in the matrix $P$ can be
generated by using Not gates, Swap gates and one or two Toffoli
gate(s). Example 1 shows this process.\\

\textbf{Example 1}: Let $n = 3, u = (0, 0, 2), s = (0, 1, 2), t =
(0, 2, 2)$. The column 2 is heterogeneous.
\[
\hspace{-1cm}
\left[\begin{array}{c}u\\ s\\ t\end{array}\right]=
\left[\begin{array}{c}0,0,2\\ 0,1,2\\ 0,2,2\end{array}\right]
{E_{1,2}\atop
\rightarrow} N_2\star (N_3)^2 \left[\begin{array}{c}0,1,1\\ 1,1,1\\
2,1,1\end{array}\right] {T\atop \rightarrow}
\left[\begin{array}{c}1,1,1\\ 2,1,1\\ 0,1,1\end{array}\right]
\]
\[
\hspace{-0.6cm}
{(N_3)^{-2}\star (N_2)^{-1}\atop \rightarrow}
\left[\begin{array}{c}1,0,2\\ 2,0,2\\ 0,0,2\end{array}\right]
{(E_{1,2})^{-1}\atop\rightarrow} \left[\begin{array}{c}0,1,2\\
0,2,2\\ 0,0,2\end{array}\right] = \left[\begin{array}{c}s\\ t\\
u\end{array}\right]
\]
Therefore,
\[
\hspace{-1cm}
\begin{array}{r@{=}l}
(u, s, t)&E_{1,2}\star N_2\star N_3\star N_3\star T\star (N_3\star N_3)^{-1}\star (N_2)^{-1}\star (E_{1,2})^{-1}\\
     &E_{1,2}\star N_2\star N_3\star N_3\star T\star N_3\star N_2\star N_2\star E_{1,2}.
\end{array}
\]
We use notation $(N_3^{-1}) (N_3^{-1}) = (N_3)^{-2}$.

In fact, at the end of step 3, we can write a generating expression
of $(u, s, t)$ as a product of the Swap gates, Not gates, and
Toffoli gates without performing step 4. We perform step 4 in
Example 1 just to show that this process is correct.\\

\textbf{Case 2}: There are two heterogeneous columns among $u$, $s$
and $t$.
\begin{enumerate}
\item Using Swap gates, we can exchange columns such that the
first and second columns are heterogeneous and the number of
different elements in the first column is no more than that in the
second column.
\item Using Not gates, set all the elements in the
homogeneous columns as values 1. \item Using Swap, Not, and Toffoli
gates, set the elements of the second columns as value 1. We have
the following three subcases:
\begin{itemize}

\item Subcase 1: There are two different elements in the first
column and three different elements in the second column. Without
loss of generality, we assume $u_1 = s_1 \neq t_1$. Consider $t_2$.
If $t_2\neq 1$, use $N_2$ (if $t_2 = 0$) or $N_2\star N_2$ (if $t_2
= 2$) to interchange $t_2$ to 1. Then use T (if $t_1\bigoplus _3 1 =
u_1$) or $T\star T$ (if $t_1\bigoplus _3 2 = u_1$) to interchange
$t_1$ to $u_1$. If $u_1 = s_1 = t_1\neq 1$, use $N_2$ or $N_2\star
N_2$ to make the elements in column 1 be 1s. Finally, exchange
columns 1 and 2. As a result, the elements in the first column are
different and the elements of other elements in $P$ are all 1s.

\item Subcase 2: There are two different elements in the first
column and the second column, respectively.  Without loss of
generality, we assume $u_2 = s_2\neq t_2$. Then $u_1\neq  s_1$. We
use Not gate(s) to change $u_2$ and $s_2$ to 1s if they are not 1s.
Then use $T$ or $T\star T$ to change $u_1$ and $s_1$ such that the
elements in the first column are different with each other. Finally,
exchange columns 1 and 2. Then, the resulting matrix $P$ becomes the
subcase 1.

\item Subcase 3: There are three different elements in the first
column and the second column, respectively. Without loss of
generality, we assume $u_2 = 1$. After using $T$, we change $u_1$ to
$s_1$ or $t_1$. Then, the resulting matrix $P$ becomes the subcase
1. For instance,
\[
\hspace{-1cm}
\left[\begin{array}{c}u\\ s\\ t\end{array}\right]=
\left[\begin{array}{c}0,2,1\\ 1,0,1\\ 2,1,1\end{array}\right]
{T\atop \rightarrow} \left[\begin{array}{c}0,2,1\\ 1,0,1\\
0,1,1\end{array}\right] \textrm{(This is subcase 1)}.
\]
\end{itemize}

\item Using Toffoli gate $T$ or $T\star T$ to change the order of the
first three vectors as expected (we can reorder the rows $r_1, r_2,
r_3$ to $r_2, r_3, r_1$).

\item Finally, using the inverse of these Not gates, Swap gates
and Toffoli gates in the steps 3, 2, and 1 to recover these changed
digital numbers, we obtain the 3-cycle $(u, s, t)$.
\end{enumerate}

The action of the $5^{th}$ step is:
\[
P^{(4)}{step 5\atop \rightarrow}P^{(5)}= \left[\begin{array}{c}s\\
t\\ u
\end{array}\right], Q^{(4)}{step 5\atop \rightarrow}Q^{(5)}=Q.
\]
Example 2 shows the process executed in case 2.\\

\textbf{Example 2}: Let $n = 3$, $u = (0,0,1)$, $s = (0,0,2)$, $t =
(1,0,1)$.
\[
\hspace{-2cm} \left[\begin{array}{c}u\\ s\\ t \end{array}\right]=
\left[\begin{array}{c}0,0,1\\ 0,0,2\\ 1,0,1\end{array}\right]
{E_{2,3}\atop \rightarrow} \left[\begin{array}{c}0,1,0\\ 0,2,0\\
1,1,0\end{array}\right] {N_3\atop\rightarrow}
\left[\begin{array}{c}0,1,1\\ 0,2,1\\ 1,1,1\end{array}\right]
\textrm{(Step l and 2)}
\]
\[
\hspace{-1cm}
{T\atop \rightarrow} \left[\begin{array}{c}1,1,1\\ 0,2,1\\
2,1,1\end{array}\right] {E_{1,2}\atop\rightarrow}
\left[\begin{array}{c}1,1,1\\ 2,0,1\\ 1,2,1\end{array}\right]
\left(\begin{array}{c}\textrm{Now it}\\ \textrm{becomes}\\
\textrm{subcase 1}\end{array}\right) \textrm{(End step 3)}
\]
\[
\hspace{-1cm}
{T\star T\atop\rightarrow} \left[\begin{array}{c}1,1,1\\ 0,1,1\\
2,1,1\end{array}\right] \textrm{(End step 4)} {(E_{1,2}^{-1})\star
(T\star T)^{-1}\star (N_2)^{-1}\atop\rightarrow} \left[\begin{array}{c}2,0,1\\
1,2,1\\ 1,1,1\end{array}\right]
\]
\[
\hspace{-1cm}
{(E_{1,2})^{-1}\star T^{-1}\atop\rightarrow} \left[\begin{array}{c}0,2,1\\ 1,1,1\\
0,1,1\end{array}\right] {(N_3)^{-1}\star
(E_{2,3})^{-1}\atop\rightarrow} \left[\begin{array}{c}0,0,2\\
1,0,1\\ 0,0,1\end{array}\right]= \left[\begin{array}{c}s\\ t\\
u\end{array}\right]
\]
Therefore,
\[
\hspace{-1cm}
\begin{array}{rcl}
(u, s, t)&=&  E_{2,3}\star N_3\star T\star E_{1,2}\star N_2\star T\star T\star E_{1,2}\star (T\star T)\star (E_{1,2})^{-1}\\
     &&  \star (T\star T)^{-1}\star (N_2)^{-1}\star (E_{1,2})^{-1}
     \star (T)^{-1}\star (N_3)^{-1}\star (E_{2,3})^{-1}\\
     &=&E_{2,3}\star N_3\star T\star E_{1,2}\star N_2\star T\star T\star E_{1,2}\star (T\star T)\star E_{1,2}\\
     &&\star T\star N_2\star N_2\star E_{1,2}\star T\star T\star N_3\star N_3\star E_{2,3}.
\end{array}
\]
In fact, after step 4, we can write a generating expression of $(u,
s, t)$ as a product of the Swap gates, Not gates, and Toffoli gates
without executing step 5. We perform step 5 in Example 2 just to
show that the process is correct.\\

\textbf{Note}: After finishing the whole process in case 1 and 2,
the remaining 27-3 = 24 rows are not affected by the string of
gates. And in the process, we can find the realization without
considering these 24 rows. Thus, we only act these gates on the
three rows $u$,
$s$ and $t$.\\

\textbf{Case 3}: There are more than two different bits among $u$,
$s$ and $t$.

Similar to the binary reflective Gray code ~\cite{Sandige:DDE:02},
we can also reflectively encode the ternary vectors in an order
$x_1, x_2,\ldots , x_m$, where $m = 3^n$ such that there is only one
bit different between two vectors $x_i$ and $x_{i+1}$, for $1\leq
i\leq m-1$. Therefore, we can find $i < j < k$, such that $x_i$,
$x_j$, and $x_k$ are a permutation of $u$, $s$, and $t$,
respectively. Namely, $(u, s, t) = (x_i, x_j, x_k)$ or $(u, s, t) =
(x_i, x_j, x_k)^2$.

There are at most two different bits among $x_h, x_{h+1}, x_{h+2}$,
for $1\leq h\leq m-2$. According to case 1 and case 2, the 3-cycle
($x_h, x_{h+1}, x_{h+2}$) can be generated by Swap, Not, and Toffoli
gates. Thus, according to Lemma~\ref{evenodd}, the 3-cycle ($x_i,
x_j, x_k$) can be generated by Swap, Not, and Toffoli gates. As a
result, $(u, s, t)$ can be generated by Swap, Not, and Toffoli
gates. \hfill{$Q.E.D$}

\end{document}